# Semantic Web Content Accessibility Guidelines for Current Research Information Systems (CRIS) and Web content developers of research relevant information at the universities and research institutions

# Summary

The most exciting challenge for CRIS is to create a service for research information which should be wide-spread, distributed and actual like Google, but at the same time structured, trusted, with a complex search and navigation similar to today's CRIS application. The core technology for such a "new" CRIS is the semantic web technology to integrate database contents with HTML/XML web pages for being provided to the research interested public. One (at the moment the best) possible way is to use RDF (Resource Description Framework) which is also recommended by the W3 consortium.

However RDF alone is not sufficient, there is a need for a European cooperation based on the EuroCRIS platform to agree on topics based on CERIF 2000 expressed in RDF - and also a need to develop the tools to create RDF files out of databases and support the publishing researcher to provide his information in the agreed format in the semantic web. That is the reason to for this publication. It should be a step by step help for all persons and organization units which have research information to be provided in the web and want to use the latest technology.

# RDF usage in AURIS-MM and CRIS

The main purpose of creating an RDF description for research information resources is the integration of research data into the semantic web. This integration allows to implement better access to data, based on their meaning, more rich search, collection, information navigation facilities for research data. The usage of Semantic Web technologies, particularly RDF for research data allows to implement a metasystem, which can accumulate data from distributed and heterogeneous data sources, such as different information systems, web pages of researchers and institutions. The usage of Semantic Web technologies allows to implement such a system which is wide-spread, distributed and actual as Google, but structured, trusted, with complex search and navigation facilities such as applications. So the implementation of a metasystem, which covers most research data of Austria or Europe from different institutions, not requiring to put or enter them into one central node, is the first application of the Semantic Web technologies for CRIS.

XML or SGML technologies are not sufficient for the development of such a distributed system. Research data of different systems have different meanings. An all embracing schema cannot be developed. Each system must publish its own data and describe their meaning. RDF technologies allow to describe the meaning of data and it is a main rationale behind usage of them.

General research data must be (following http://www.w3.org/1999/04/WebData)

Evolution and partial understanding. The distributed CRIS must permit distributed communities to work independently to increase the Web of understanding, adding new information without insisting that the old must be modified. This approach allows the communities to resolve ambiguities and clarify inconsistencies over time while taking the maximum advantage of the wealth of backgrounds and abilities reachable through the Web. Therefore the distributed must be based on a facility that can expand as human understanding expands. This facility must be able to capture information that links independent representations of overlapping areas of knowledge.

Global universality and local constraint. In an object-oriented model or in XML schemas, the variables in the object are declared when the object type is declared. But distributed CRIS should be able to represent these constrained model, but, as with link consistency, we must relax absolute constraints to achieve scalability; when an object is exported to the Web, the "anything can say anything about anything" rule allows assertions to be made about the object expressing things which were not foreseen in the original definition of that object.

The mechanism adopted in RDF to manage the expression of constraints is to make all objects, all relationships, all types, and even all assertions be "first class objects" on the Web. That is, they have their own URLs and are not constrained in the fundamental level to be combined in any particular way. By giving first class identifiers to types, relationships, and assertions we allow the Semantic Web to make assertions about itself.

The second application of RDF technologies is the implementation of a data exchange solution for different CRIS for data sharing. RDF encoding of data maybe used for dissemination research results, developing the system with an almost complete research information.

The third application of RDF technologies is the development of metasearch and data gathering facilities. Putting RDF descriptions of projects, persons, organization units and publications into the web or including RDF descriptions into web pages makes it possible to develop complex metasearch engines and gather data from web pages. Actually it is a more light-weight variant of the first solution (semantic web).
Furthermore there is more detailed description of these three possible applications of RDF.

## 1 Semantic web.

## Why Semantic Web?

Why apply the semantic web technology for research information?

The web today is based on links from one document to anything that is accessible in the Internet. From HTML pages only humans can read and understand why a link is set. The semantic web provides "links with meaning". Let us see a simple example (the following question) which could be possible in the semantic web:

"What is the special field of research of the rector of theVienna University of Technology"

At the moment it is necessary to search / read web pages as follows:
- search for Vienna University of Technology ---> http://www.tuwien.ac.at
- read the found web page and type into search input field: "Rektor" (you must know that the English "rector" is in German "Rektor")
  - Ah, you get a "White pages" web page which provides a link "Rektor, Buero des Rektors, Sonstige Einrichtungen"
    !Click on it!
  - Ah, you get another "White pages" web page and again a link
"Zustaendige Person Peter Skalicky (E137), Personal,
Technische Universitaet Wien, Österreich"
!Click on it!
  - Ah, you get another "White pages" web page with the title "Peter Skalicky (E137)" and again a link under the attribute "Geschäftsbereich"
    - Angewandte und Technische Physik
    - Buero des Rektors
    (you must know that "Geschäftsbereich" is a general German word for "field of activities" and you must know that "field of activities" means something 70% similar like "field of research" ... and you must know that "Angewandte und Technische Physik" means in English "Applied and technological physics"

You got it...

By using the semantic web technology (RDF) in the future you will be able to use heuristic logics and the triples used in RDF:

1. search (Vienna University of Technology, URL, x) yields http://www.tuwien.ac.at
2. search (http://www.tuwien.ac.at, rector, x) yields a nil result
   search (rector, equivalent, x) yields x=Rektor
3. search (http://www.tuwien.ac.at, Rektor, x) yields the requested "White pages" web page

based on an agreement on the address http://rdf.tuwien.ac.at where the meta-info structure of the

Vienna University of Technology is stored.
RDF technology makes it possible to use in information retrieval operations semantic definitions of elements, types, vocabulary items, element's attributes and relations.

One of the possible applications of RDF is the development of Semantic Web solutions. These solutions can present distributed knowledge about research in one uniform network, which can be used for presentations, visualizations, search and analysis of research data. Such a network maybe created not as a database, a number of records on one computer but as a distributed set of information resources. Semantic web technologies allow to understand the meaning of such resources, their attributes, and values of the attributes, they can ask complex queries over all the information space, check if the information can be trusted and so on.

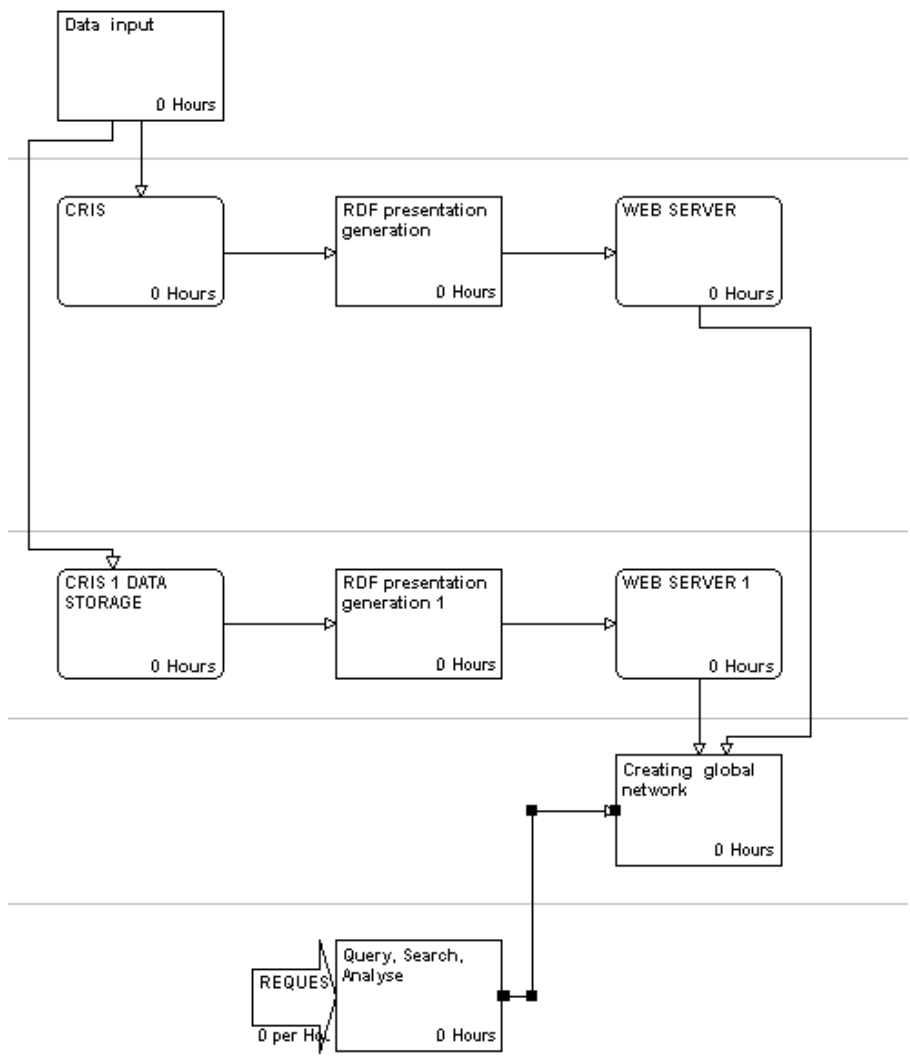

**Fig. 1. Semantic web facilities over RDF descriptions.**
**Description of Figure 1.**
**Step 1. Researchers input their data into RDF Semantic Web compatible CRIS. A number of researchers input their data  into CRIS, other researchers input research data into CRIS1. To make data accessible to all users of distributed CRIS it is enough to publish data into one CRIS only; they become accessible to users of all other CRIS. Researchers can publish their data into CRIS using web forms of CRIS, RDF loaders of CRIS or just put their metadata in**

**CERIF-RDF on own pages.**
**Step 2. CRIS, using RDF presentation generator generates RDF description of the data in CERIF-RDF format. CRIS can use own RDF format of data and even ontology, but to make data accessible by CRIS users in trusted and proper way, CRIS must provide RDF Schema description and ontology description in CERIF-RDF terms.**
**Step 3. RDF description of data is published at web site of CRIS. So any agent or search engine, or even browser can access and use CRIS data.**
**Step 4. Agents, presentation services and metasearch engines use CRIS CERIF-RDF data of multiple CRIS to create a global network of CRIS data. Technologies such as Redland can be used.**

Global knowledge systems based on RDF can be developed by such tools as the Redland RDF Application Framework.
The Redland RDF Application framework (http://www.redland.opensource.ac.uk/) - a set of API to build RDF applications.

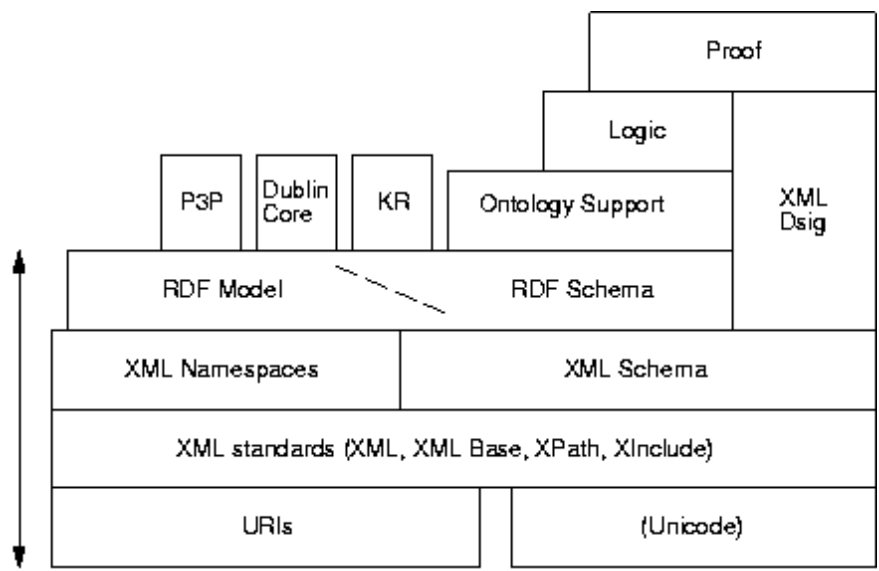

**Redland RDF building blocks (from Redland guidelines).**

## 2 Data exchange.

There is a strong need to integrate and exchange data between information systems. Such a solution would always be required despite the Semantic Web, "metasystem" and other solutions. The main advantages of data exchange are:

- data can be entered by the researcher only once into one CRIS, then the researcher's data will appear in all systems
- data imported into the information system can be controlled by the information system for authorized access, change
- they can be put into workflow processes, controlling them and acting on their changes
- version control can be applied for storing all versions of data and their changes
- tools for data analysis needed for research management of can be applied to imported data
- data from different systems can be linked to each other to create a more complete presentation

of information about the research

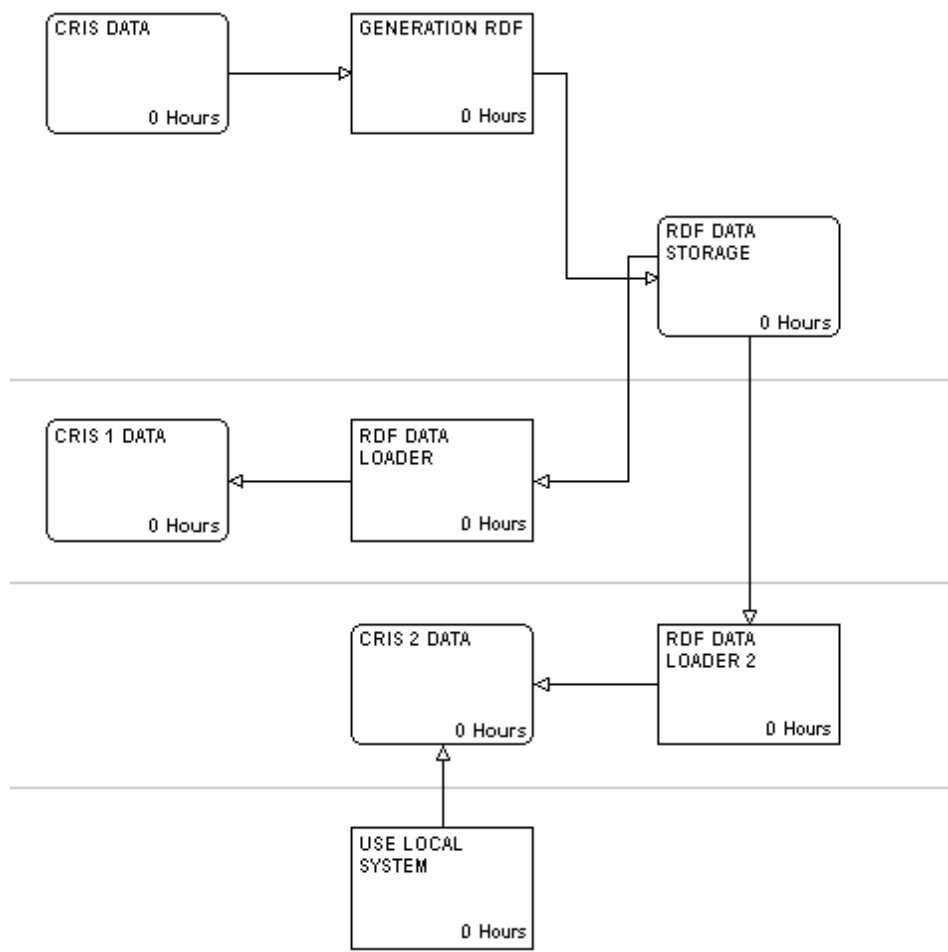

**Fig. 2. Data exchange**
**Description of figure 2.**
**Step 1. CRIS 1, using RDF generator generates RDF presentation of its data.**
**Step 2. Data is published on the web. RDF Schema of data, ontology, thesaurus definition must also be published ifthey are different from CERIF-RDF.   Also a RDF Site Summary description of data can be published.**
**Step 3. Any CRIS, which needs data of CRIS 1, having knowledge about CRIS 1 data storage load RDF data presentation of CRIS 1 data or part of it.**
**Step 4. Using RDF loader tools, any CRIS which need data of CRIS 1, load CRIS 1 data into own database.**
**Step 5. Data of CRIS 1 are accessible to others CRIS  users as it was native data for that CRIS.**

The RDF technology solution has a number of advantages to theXML solution
- semantic of new elements and attributes can be described
- RDF "anything can say anything about anything" paradigm
- information not only about research information, but also about statements about research information can be exchanged
- predefined collection properties can be used

## 3 Data gathering facilities.

One of the aims of the CRIS development is gathering research data into CRIS.  Researchers do not always  have time and will to put data into CRIS. Also some researchers do not know about the benefits of putting data into CRIS. If there are a number of CRIS, entering research data in them all can be a real benefit for the researcher but it is a hard and time-consuming task. One possible solution to this problem is to enable the researcher  to put data somewhere one time and take data for all CRIS from that source. Such a place already exists, most researchers already publish data into the web as a set of html pages with information about projects, persons, organization units, publications.  But as usual data is putted in weak-structured way and difficult to extract from a web page by software. If the data are published inhtml, due to absence of common standards for html data publishing about research, also it is impossible to gather data from different universities orteams. One of the way is data publishing in a standardized way in RDF, including this RDF into web pages, without changes in their design. This solution makes it possible forrobots  to gather data and understand them and also  makes it possible for researchers to develop their pages as they want. But publishing data about research in RDF can also be very helpful for researchers.  XSLT tools for RDF to HTML transformation can make HTML presentation of RDF data, making HTML generation very easy and not taking a lot of time for researchers.

Publishing data in RDF by CRIS, using such tools as suggested by AURIS-MM  or exactly by researchers  makes is  possible to implement a lot of data exchange or integration solutions without any additional work. Any information system can ask already generated data. So this solution allows to integrate data of CRIS with any number of other CRIS having RDF gather data tools.

Also it allows to develop global knowledge systems based on RDF for such tools as the Redland RDF Application Framework.

The Redland RDF Application framework (http://www.redland.opensource.ac.uk/) - a set of API  to build RDF applications.

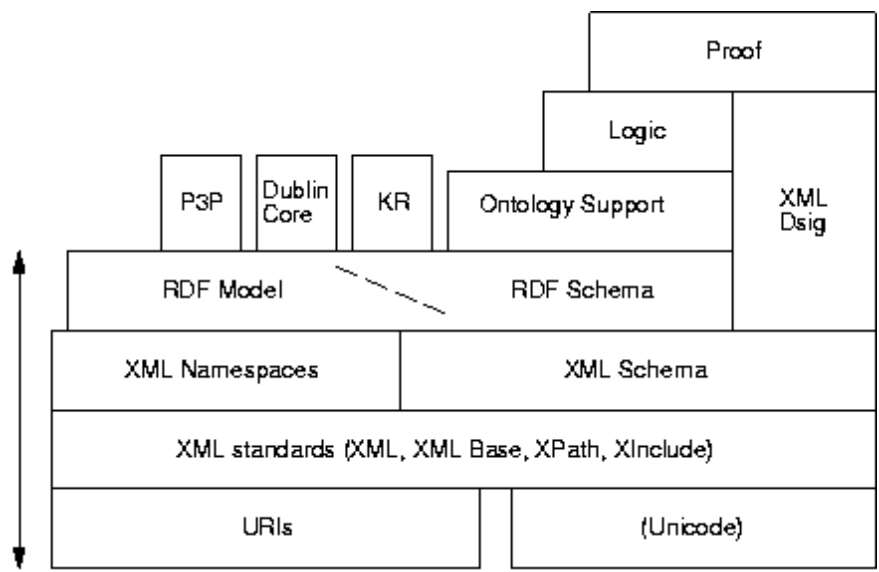

**Redland RDF building blocks (from Redland guidelines).**

RDF's essential aim is to make work easier for autonomous agents, which would refine theWeb by improving search
engines and service directories.

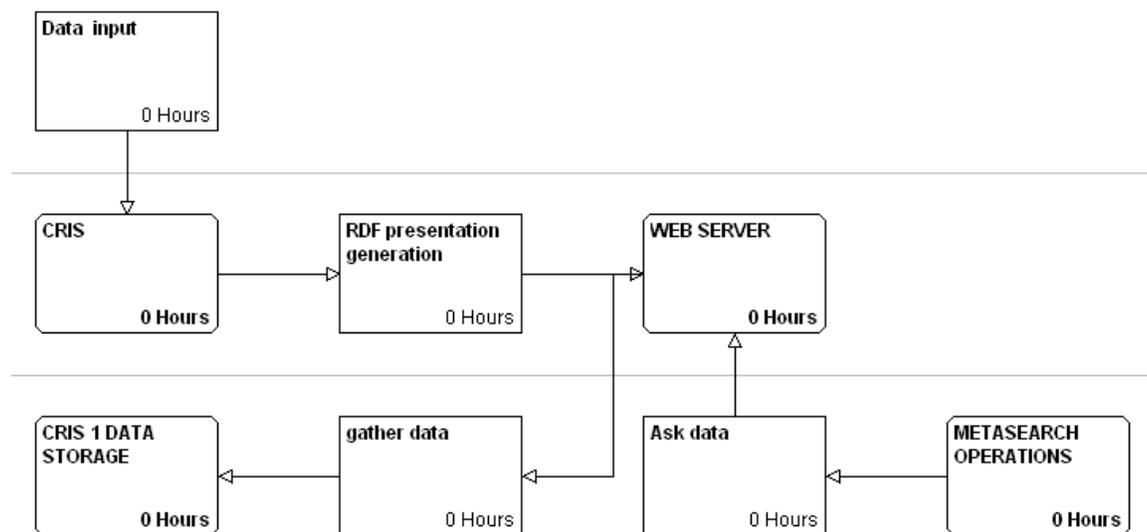

**Fig 3.**

**Diagram of RDF usage for data gathering and metasearch**

## Example of HTML file with RDF description

# Simple Examples

Our aims is to develop the **AURIS-MM network** – set of applications, relational or RDF databases and other information sources and services, serving for research in Austria. Each application or database to be a part of the AURIS-MM network must be compatible with the AURIS-MM network standards. It must publish data in a way compatible with CERIF-2000 (CERIF database from CORDIS), or CERIF-RDF (RDF Schema for CERIF from TUWIEN). Now the AURIS-MM network consist of

| Application or database | Supports |
|---|---|
| AURIS-MM prototype developed in the Extension Centre of the TUWIEN | CERIF database<br>CERIF-RDF Schemadata importing and exporting |
| AURIS database | Mapping from database to CERIF database is developed |

The CERIF format declares the structure of data and the vocabularies used. To make the AURIS-MM network easier to implement and easier to adopt by Austrian researchers, it was developed as AURIS-CERIF format, which is a part of CERIF in structure of data and in which the Austrian vocabulary is used. Data published in the AURIS-CERIF format can be accessed by Austrian users, using their specific vocabularies and by European users, using European vocabularies.

In this part there are simple examples of publishing data in a way compatible with AURIS-CERIF.

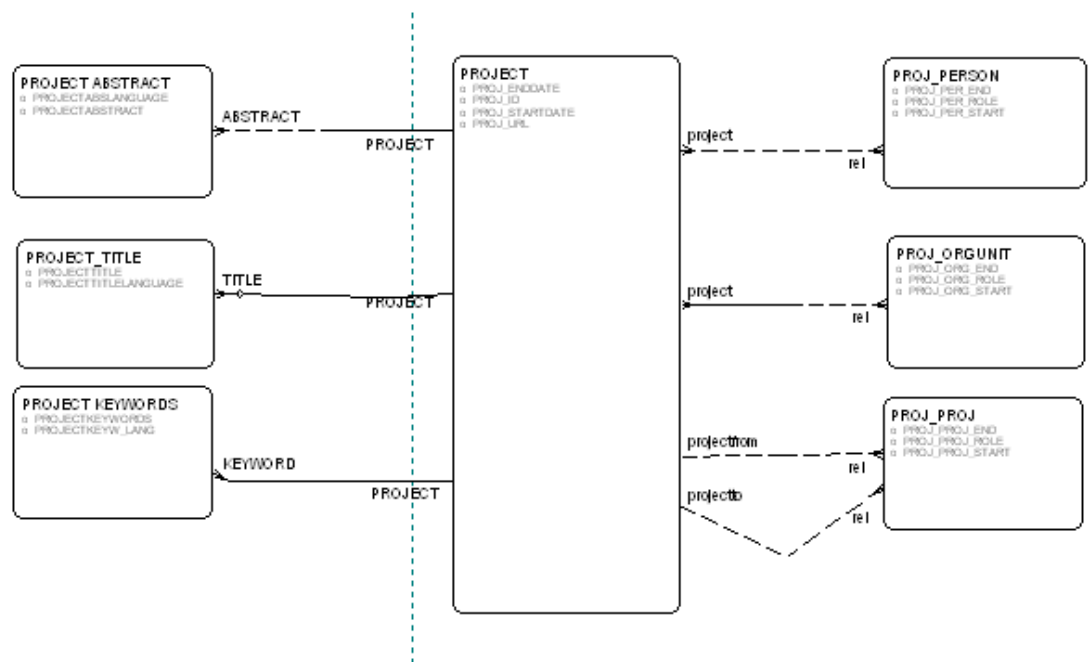

Figure L1.1.  Part of CERIF  Entity-Relationship diagram  for project

## *HTML*

The HTML language does not allow to define data structures; it is a simple language for presentation. But already at the publication stage, the data presentation can be structured in a format which makes it easy to extract data into AURIS-CERIF.

There are two different layouts of the project description html page. The first one is from the Fodok database and that page has not layout close to the CERIF model. The second page is very close to the CERIF model in layout and CERIF compatible metadata can be easily extracted from that page.

## **Presentation of information about project AURIS-MM in FODOK. Not close to CERIF.**

**Forschungsdokumentation Online**
Außeninstitut der Technischen Universität Wien

Home | Search | Update | Products | Feedback | Help | About

Technische Universität Wien

Dienstleistungseinrichtungen und Serviceinstitute

Außeninstitut

## Forschungs-Projekt F015-01-08

**Titel:** Multimediale Neugestaltung und Erweiterung von AURIS (Austrian Research Information System) zur Steigerung der Attraktivität und Benierfreundlichkeit der österreichischen Forschungsdokumentation

**Titel (englisch):** Austrian Research Information System: Multimedial Enhancement

**Leiter (Kontaktperson):**
Name: Hiedenmayer, Walter (Dipl.-Ing.)
Spezialgebiet: Forschungsdokumentation, Current research information systems
E-Mail-Adresse: walter@derpi.tuwien.ac.at
…
Name: Krieger, Andreas
Spezialgebiet: Datenverarbeitung, Programmierung, Datenbank
E-Mail-Adresse: krieger@derpi.tuwien.ac.at

**Auftraggeber / Forschungsprogramm / Programm-Akronym / Projektnummer:** BMVIT

**Beginn-Datum:** 2000-02

**Ende-Datum:** 2001-12

**Forscher:**
Name: Krieger, Andreas
Spezialgebiet: Datenverarbeitung

**Beschreibung:** Das Forschungsdokumentationssystem AURIS wurde im September 1998 von der Arbeitsgemeinschaft "Österreichische Forschungsdokumentation" fertiggestellt und steht seither für Online Recherchen im Internet frei zur Verfügung. Es sind 12 österreichische Universitäten beteiligt. Die AURIS-Datenbank umfaßt derzeit etwa 15.000 Projektbeschreibungen.
Um die außerfällige Akzeptanz von AURIS sicherzustellen, sind eine attraktive Palette von Recherche-Möglichkeiten, die Einbeziehung multimedialer Elemente bei der Präsentation von Forschungs-Leistungen, eine Kontrolle und Sicherstellung der Datenqualität und die systematische Einbindung weiterer vorhandener Forschungs-Informationen mittels einer Internet-Suchmaschine geplant. Neuartige Methoden zur Identifizierung, Auswahl, Kontrolle und Einbeziehung von Web-Sites in das Suchbereich von AURIS sowie die optionale Darstellung der österreichischen Forschungsinformationen als …

In mehreren Arbeitsgruppen werden Lösungen zu Problemstellungen zu Bilder-, Video-Formaten, Richtlinien für technische Hilfen bei der Kennzeichnung und Einbeziehung multimedialer Forschungs-Informationen, Fragen des Qualitätsmanagements, der Evaluierung, linguistische Methoden, Visualisierungsmöglichkeiten in Form von virtuellen Forschungslandschaften, ein Konzept und die Erstellung einer CD-ROM (bzw. DVD), Marketing und Online-Recht behandelt. Bei der Arbeit aller Arbeitsgruppen (genaueres in 3.1.5) wird das Internet als Informations- und Kommunikationsbasis (Nutzung bereits vorhandener Informationen im Web, Online-Diskussionen, Telekonferenzen) verwendet. Es wird eine gesamtösterreichische Lösung angestrebt. Die Online-Diskussionsforen sollen prinzipiell auch für externe und ausländische Interessenten offen stehen. Der jeweils aktuelle Stand des Projektes wird für die interessierte Öffentlichkeit online im Web dokumentiert werden (http://arge.tuwien.ac.at).
Die Arbeitsgruppen:
AG1 Multimediale Elemente - Einbringung
AG2 Online-Tool-Kit&Suchmaschine - Konzept&Realisierung
AG3 Qualitätsmanagement
AG4 Linguistik und Visualisierung
AG5 CD-ROM (bzw. DVD) - Konzept und Produktion
AG6 Marketing, Web-Design und Realisierung der Services
AG7 Rechtsfragen
AG8 Projektkoordination und -management

**Beschreibung (englisch):** The research information system AURIS http://www.auris.ac.at was implemented in September 1998 by the "Arge Österreichische Forschungsdokumentation" (the joint committee Austrian Research Documentation). The system is based on an Oracle database with currently more than 15.000 documents on research units and research activities. It is available online for database query. To ensure the acceptance of the AURIS database by the public, the project AURIS-MM will provide the following extensions to the AURIS service:
1. Enlarging the search run:
Making documents from AURIS and further Austrian research databases like AUFDAT, FWF or BM searchable by users all together in only one single search run (highly structured web documents) Automated identification, selection, quality control and inclusion of web sites into the search area of AURIS-MM using the Dublin Core Meta Data Set (web documents with only a general structure according the Dublin Core Standard for Metadata). Indexing the contents of all research relevant web sites in Austria (universities and other research institutions) by an internet search engine to be searchable in one search run (almost unstructured research information).
2. Additional services:
Additional services like email notification, creating on-line brochures, building and visualization of search results sets, browsing of AURIS documents is part of the project. The option to navigate Austrian research information with help of virtual landscapes.. Multimedia data acquisition: A publishing toolkit with FAQs for digitizing, encoding multimedia elements for the web usage) for on-line multimedia elements which can be uploaded to web pages, multimedia servers or into database. quality control for the data and the service Registering research relevant web documents also for international Internet search engines
The following working groups concentrate on requirements in these areas: AG1 video/audio formats to be used AG2 specification and installation of a content management system and internet search engine AG3 questions of quality management and evaluation AG4 linguistic methods and visualization by way of virtual landscapes of research AG5 concept and realization of a CD-ROM about AURIS AG6 marketing and special services AG7 rights in online issues AG8 project management
All working groups use the internet as basis for communication and information collection (find existing information on the web, online discussion, teleconferencing). An Austrian-wide solution is achieved. The online discussions are open to any interested audience. The current status of the project is documented at http://arge.tuwien.ac.at/.

**Schlagwörter:** Österreichische Forschungsdokumentation; Multimedia in der Forschungsdokumentation; AURIS Multimedia

**Schlagwörter (englisch):** Austrian research information; multimedial AURIS; multimedia in research documentation

**ÖSTAT Klassifikation:** 1109 Dokumentations- und Datenverarbeitung; 1138 Informationssysteme (5937); 1105 Computer Software

**Ziel des Projektes (useful for):**
Anwendungsgebiet (englisch): national and international research cooperation and facilitation
Marktfähigkeit der Projektergebnisse für dieses Anwendungsgebiet: Test/ Pilotanwendung/ Prototyp/ Demonstration

**Durchschnittliche Anzahl der vollbeschäftigten Forscher (Intensität):** 10

**Personen-Monate (Gesamt-Arbeits-Aufwand):** 150

**Kooperationen mit Institutionen:**

**Presentation of information about the AURIS-MM project, published in HTML according to AURIS-CERIF. Close to CERIF.**

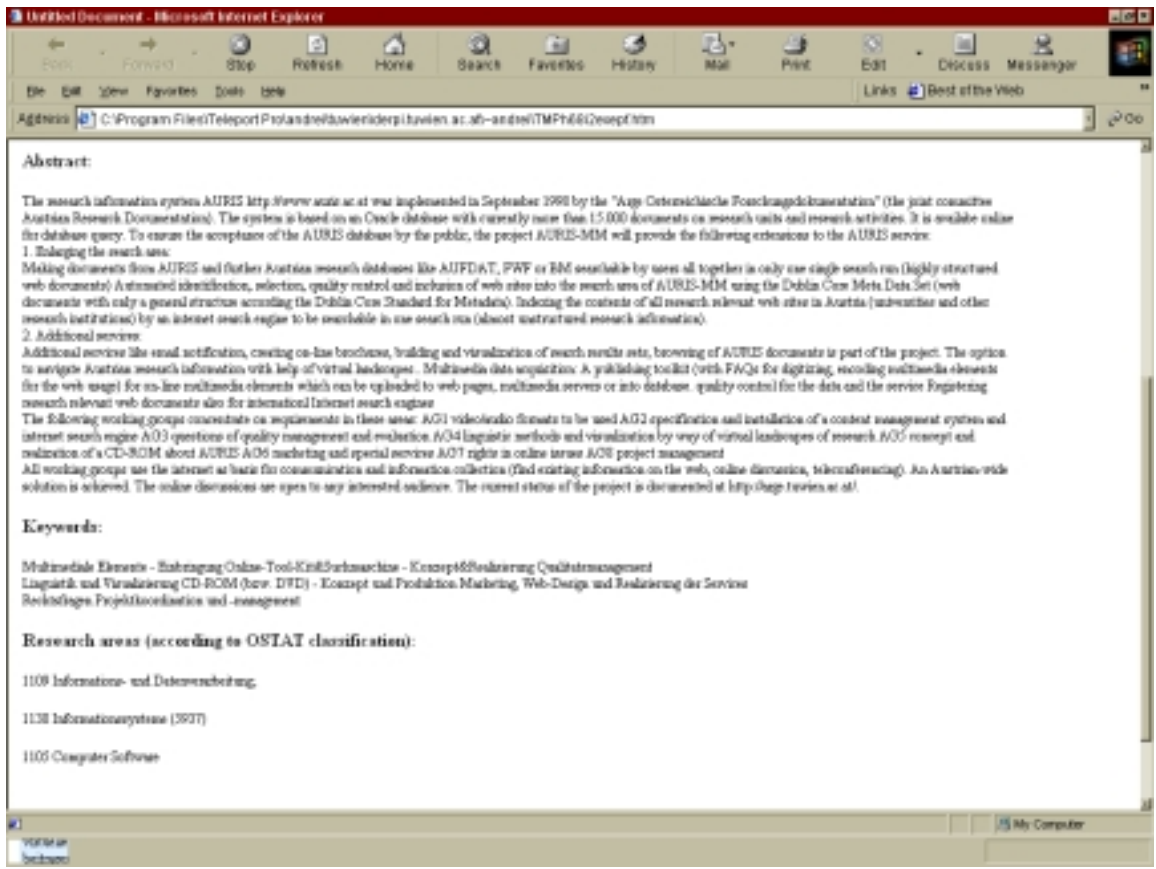

### *RDF presentation of AURIS-CERIF data*

Means of RDF elements, properties and instructions how to generate them are written in step-by-step part of this guidelines.

### Example of RDF project description for a project.

```
<rdf:RDF xmlns:rdf="http://www.w3.org/1999/02/22-rdf-syntax-ns#"
xmlns:rdfs="http://www.w3.org/2000/01/rdf-schema#"
xmlns:cerif="http://derpi.tuwien.ac.at/~andrei/cerif-rdf#">
 <cerif:project ID="E015-01-08">
  <cerif:proj_status>
   Execution
  </cerif:proj_status>
  <cerif:proj_startdate>
   02.2000
  </cerif:proj_startdate>
  <cerif:proj_enddate>
   12.2001
  </cerif:proj_enddate>
  <cerif:proj_uri>
   http://arge.tuwien.ac.at
  </cerif:proj_uri>
  <cerif:proj_prizeaward>
  </cerif:proj_prizeaward>
  <cerif:project-abstracts>
   <rdf:Bag>
    <rdf:li>
     <cerif:Project-abstract>
      <cerif:proj_abs_language>
       en
```

```
        </cerif:proj_abs_language>
        <cerif:proj_abs_trans_type>
         H
        </cerif:proj_abs_trans_type>
        <cerif:proj_abstract>
        The research information system AURIS http://www.auris.ac.at was implemented in
September 1998 by the "Arge Osterreichische Forschungsdokumentation" (the joint committee
Austrian Research Documentation). The system is based on an Oracle database wit
        h currently more than 15.000 documents on research units and research activities. It
is availabe online for database query. To ensure the acceptance of the AURIS database by the
public, the project AURIS-MM will provide the following extensions
        to t      he AURIS service:
        </cerif:proj_abstract>
       </cerif:Project-abstract>
      </rdf:li>
      <rdf:li>
       <cerif:Project-abstract>
        <cerif:proj_abs_language>
         de
        </cerif:proj_abs_language>
        <cerif:proj_abs_trans_type>
         O
        </cerif:proj_abs_trans_type>
        <cerif:proj_abstract>
        Das Forschungsinformationssystem AURIS wurde im September 1998 von der
Arbeitsgemeinschaft "Osterreichische Forschungsdokumentation" fertiggestellt und steht
seither fur Online Recherchen im Internet frei zur Verfugung. Es sind 12 osterreichische Un
        iversitaten beteiligt. Die AURIS-Datenbank umfa?t derzeit etwa 10.000
Projektbeschreibungen.
        </cerif:proj_abstract>
       </cerif:Project-abstract>
      </rdf:li>
     </rdf:Bag>
    </cerif:project-abstracts>
    <cerif:project-titles>
     <rdf:Bag>
      <rdf:li>
       <cerif:Project-title>
        <cerif:proj_title_language>
         en
        </cerif:proj_title_language>
        <cerif:proj_title_trans_type>
         H
        </cerif:proj_title_trans_type>
        <cerif:proj_title>
         Austrian Research Information System: Multimedial Enhancement
        </cerif:proj_title>
       </cerif:Project-title>
      </rdf:li>
      <rdf:li>
       <cerif:Project-title>
        <cerif:proj_title_language>
         de
        </cerif:proj_title_language>
        <cerif:proj_title_trans_type>
         O
        </cerif:proj_title_trans_type>
        <cerif:proj_title>
         Multimediale Neugestaltung und Erweiterung von AURIS (Austrian Research Information
System) zur Steigerung der Attraktivitat und Bedienerfreundlichkeit der osterreichischen
Forschungsdokumentation
        </cerif:proj_title>
       </cerif:Project-title>
      </rdf:li>
     </rdf:Bag>
    </cerif:project-titles>
    <cerif:project-keywords>
```

```
  <rdf:Bag>
   <rdf:li>
   </rdf:li>
   <rdf:li>
   </rdf:li>
  </rdf:Bag>
 </cerif:project-keywords>
</cerif:project>
</rdf:RDF>
```

## Example of RDF project description for a person.

```
<rdf:RDF xmlns:rdf="http://www.w3.org/1999/02/22-rdf-syntax-ns#"
xmlns:rdfs="http://www.w3.org/2000/01/rdf-schema#">
 <cerif.person ID="273">
  <cerif.person.per_family_names>Niedermayer</cerif.person.per_family_names>
  <cerif.person.per_first_names>Walter</cerif.person.per_first_names>
  <cerif.person.per_sex>M</cerif.person.per_sex>
  <cerif.person.per_prize_awards></cerif.person.per_prize_awards>
  <cerif.person.per_uri>http://derpi.tuwien.ac.at/~walter/index-e.html</cerif.person.per_uri>
  <cerif.person.expert_skills>
   <rdf:Bag>
    <rdf:li>
     <cerif.person.expert_skill>
      <cerif.person.es.role/>
      <cerif.person.es.id>Multimedia</cerif.person.es.id>
     </cerif.person.expert_skill>
    </rdf:li>
    <rdf:li>
     <cerif.person.expert_skill>
      <cerif.person.es.role/>
      <cerif.person.es.id>CRIS</cerif.person.es.id>
     </cerif.person.expert_skill>
    </rdf:li>
   </rdf:Bag>
  </cerif.person.expert_skills>
  <cerif.person.contacts>
   <rdf:Bag>
    <rdf:li>
     <cerif.contact>
      <cerif.contact.telephone/>
      <cerif.contact.email>walter@derpi.tuwien.ac.at</cerif.contact.email>
      <cerif.contact.uri>http://derpi.tuwien.ac.at/~walter/index-e.html</cerif.contact.uri>
     </cerif.contact>
    </rdf:li>
   </rdf:Bag>
  </cerif.person.contacts>
 </cerif.person>
</rdf:RDF>
```

## Example of RDF project description for an organization unit.

```
<cerif.orgunit ID="TUWIEN.AUSENINSTITUT">
  <cerif:orgunit.org_acronym/>
  <cerif:orgunit.org_prizeaward/>
  <cerif:orgunit.org_url/>
  <cerif:orgunit.orgunit_names>
   <rdf:Bag>
    <rdf:li>
     <cerif:orgunit.orgunit_name>
      <cerif:orgunit.oun.language>de</cerif:orgunit.oun.language>
      <cerif:orgunit.oun.translation>O</cerif:orgunit.oun.translation>
      <cerif:orgunit.oun.name>Auseninstitut</cerif:orgunit.oun.name>
     </cerif:orgunit.orgunit_name>
    </rdf:li>
```

```
   </rdf:Bag>
  </cerif:orgunit.orgunit_names>
  <cerif:orgunit.ou_ou_relations>
   <rdf:Bag>
    <rdf:li>
     <cerif:orgunit.ou_ou_relation>
      <cerif:orgunit.ou_ou_r.orgunit resouce=" TUWIEN"/>
       <cerif:orgunit.ou_ou_r.role>parent</cerif:orgunit.ou_ou_r.role>
     </cerif:orgunit.ou_ou_relation>
    </rdf:li>
   </rdf:Bag>
  </cerif:orgunit.ou_ou_relations>
  <cerif:orgunit.expert_skills>
   <rdf:Bag>
    <rdf:li>
     <cerif:orgunit.expert_skill>
      <cerif.orgunit.es.role>
      </cerif.orgunit.es.role>
      <cerif.orgunit.es.skill>CRIS-Current Research Information
System</cerif.orgunit.es.skill>
     </cerif:orgunit.expert_skill>
    </rdf:li>
   </rdf:Bag>
  </cerif:orgunit.expert_skills>
 </cerif.orgunit>
```

# Guidelines. Step-by-step instructions.

How to send your data to AURIS-MM:

    **Step 1.** Define which data must be sent to AURIS-MM
    **Step 2.** Encode data into RDF
    **Step 3.** Write RDF data into files
    **Step 4.** Send files to AURIS-MM

## *Step 1 Define which data must be sent*

AURIS-MM is a CERIF based database. It stores and provides access only to CERIF types of objects. AURIS-MM can be different in data types or in the structure of data from university or institute databases. AURIS-MM can load and understand only CERIF information. So an institution or university database cansend only CERIF information to AURIS-MM. Detailed guidelines how to do it are provided in the next paragraphs. But in plans for data exchange and development of the Semantic Web, it should be counted that AURIS-MM understands information about

1)    organization units: for AURIS-MM organization units are universities, faculties, institutes, laboratories, libraries, institutions of the Austrian Academy of Science. Organization units are all organizations which are doing research, education or are involved in a research process.
2)    Publications: publications are books, articles, papers from proceedings of conferences. Only research publications are accepted by AURIS-MM.
3)    Persons: persons are researchers, involved in research, writing research articles, involved in research projects.
4)    projects,
5)    facilities,
6)   services,
7)    patents,
8)    and other CERIF entities.

AURIS-MM is a public database. All information stored in AURIS-MM can be seen by any user. So information which is secret must not be sent to AURIS-MM.

## Step 2. Encode data into RDF

To make the data exchange solution more flexible and easy to implement the CERIF RDF schema is subdivided on a number of layers. Each layer consists of a number of entities and the relation between them. In solution implementing one of the layers must exchange data described only by this layer's schema.

The layer structure can be used for dividing the complete data exchange solution on a number of steps – for implementation and testing.

## Layer 1. Projects and organization units.

The connects of this layer are information about projects, organization units and relations between them. Information about research projects is most important for researchers information types of CRIS, as was noticed in a number of articles at the CRIS conferences. The most systems in Europe and particularly in Austria store data about projects, so information about projects is a share of different CRIS. The same is the case withorganizations, because research is being done by research organizations, as usually projects are being developed by organizations (departments, labs, institutions). And to provide more detailed information about projects we need to describe which organizations take part in this project and how.

In this layer the focus is on a project, so information about organizations is very basic, mainly for the description of a project.

The ER diagram for data exchange is shown at pic L1.1. This diagram describes the ER model for CERIF.

A project is described by entity projects. There is exactly one instance of project class for each project. This entity describes main and language independent characteristics of the projects – date of start and end, URI, prize and awards, status of the project.

Level 1 Excerpt of RDF schema for CERIF. To describe a project you must

| 1 define cerif:project entity | Look at<br><cerif:project.project ID="E015-01-08"><br></cerif:project.project> |
|---|---|
| 2 assign ID attribute to this entity. It is mandatory. ID attribute maybe taken from your database primary key, in any case it must be unique in RDF document and the same for the same project in different RDF documents | <cerif:project.project ID="E015-01-08"><br></cerif:project.project> |
| 3 create cerif:proj_status element as a subelement of cerif: project and assign value, which describes status of current project<br>There are values<br>Execution –<br>Accepted<br>Completed<br>Started<br>If in your system there is no description of the status of the | <cerif:project.project ID="E015-01-08"><br><br><cerif:project.Status>Execution</cerif:project.Status><br>..<br></cerif:project.project> |
| project you need to make assumption about if for CERIF-RDF | |
| 4 describe time boundaries of your project<br>To do it you must generate element cerif:proj_start_date for start date of the project and cerif:proj_end_date for end date of the project.<br>If there is no information about the start date or end date, this element can be omitted.<br>The start date is a date when the project execution began, it is not the date of acceptance or approval.<br>The end date is the date when the project is finished according to official documents.<br>The date must be provided in a DD.MM.YYYY format. If there is no information about either day or month it can be omitted. As example for June 2000, you must write 06.2000, not 00.06.2000 | `<cerif:project.project ID="E015-01-08">`<br>`  <cerif:project.Status>`<br>`  Execution`<br>`  </cerif:project.Status>`<br><br>`<cerif:project.StartDate>02.2000</cerif:project.StartDate>`<br>`<cerif:project.EndDate>12.2001</cerif:project.EndDate>`<br>`</cerif:project.project>` |
| 5 to provide information about the project URI, mainly it is the homepage of the project, you need to create an element cerif:proj_url.<br>It is an optional element, the project may have no URI, in this case the element must be omitted.<br>The value of this element must be provided in a standard URI notation | `<cerif:project.project>`<br>…<br>`<cerif:project.URI>http://arge.tuwien.ac.at</cerif:project.URI>`<br>…<br>`</cerif:project.project>` |
| 6 to provide information about prizes or awards of the project you need to create cerif:proj_prize_award element<br>All prizes and awards should be described in one string, in a semi-column separated string.<br>This value and element are optional and can be omitted. | Example 1: No prizes<br>`<cerif:project.project>`<br>…<br>`  <cerif:project.PrizeAwards>`<br>`  </cerif:project.PrizeAwards>`<br>…<br>`</cerif:project.project>`<br>Example 2: Some prizes<br>`<cerif:project.project>`<br>…<br>`  <cerif:project.PrizeAwards>Prize 1 name; award 1 name; prize 2 name; prize 3 name</cerif:project.PrizeAwards>`<br>…<br>`</cerif:project.project>` |
| 7 to provide information about the title of the project you need to create an element cerif:project-titles. It is a mandatory element, each project must have a title. You can provide a number of titles, descriptions for each translation of a title.<br><br>Having created a cerif:project-titles element, you must create a rdf:Bag subelement of it. And then for each title translation you must create a rdf:li element with a subelement cerif:project-title<br>Each cerif:project-title element is a one translation of the project title. Translation is characterized by language, translation-type. For each cerif:project-title element you must create<br><br>■ Subelement cerif:proj_title_language and its value must be code of the language of translation according to ISO 639 Codes. En for English, de for German<br>■ Subelement cerif:proj_title_transl_type which describes type of translation, O – original, H-human, M – Machine<br>■ Subelement cerif:proj_title, describing thetitle (translation of title) | `<cerif:project.project>`<br>…<br>　`<cerif:project.project-titles>`<br>　`<rdf:Bag>`<br>　`<rdf:li>`<br>**<!—one title translation-->**<br>　　`<cerif:project.project-title>`<br>　　`<cerif:project.lang>en</cerif:project.lang>`<br><br>`<cerif:project.translation>H</cerif:project.translation>`<br>　　`<cerif:project.title>Austrian Research Information System: Multimedial Enhancement</cerif:project.title>`<br>　　`</cerif:project.project-title>`<br>　`</rdf:li>`<br>　`<rdf:li>`<br>**<!—another title translation-->**<br>　　`<cerif:project.project-title>`<br>　　`<cerif:project.lang>de</cerif:project.lang>`<br><br>`<cerif:project.translation>O</cerif:project.translation>`<br>　　`<cerif:project.title> Multimediale Neugestaltung und Erweiterung von AURIS (Austrian Research Information System) zur Steigerung der Attraktivitat und Bedienerfreundlichkeit der osterreichischen Forschungsdokumentation</cerif:project.title>`<br>　　`</cerif:project.project-title>`<br>　`</rdf:li>`<br>　`</rdf:Bag>`<br>　`</cerif:project.project-titles>`<br>…<br>`</cerif:project.project>` |
| 7 to provide information about the abstract of the project you need to create an element cerif:project-abstracts. It is a mandatory element, each project must have an abstract. You can provide a number of abstract descriptions for each translation of abstract.<br>Having created a cerif:project-abstracts element, you must create a rdf:Bag subelement of it. And then for each abstract translation you must create a rdf:li element with a subelement cerif:project-abstract | `<cerif:project.project>`<br>…<br>　`<cerif:project.project-abstracts>`<br>　`<rdf:Bag>`<br>　`<rdf:li>`<br>　　`<cerif:project.project-abstract>`<br>　　`<cerif:project.lang>en</cerif:project.lang>`<br><br>`<cerif:project.translation>H</cerif:project.translation>`<br>　　`<cerif:project.abstract>The research information system AURIS http://www.auris.ac.at was implemented in September 1998 by the "Arge Osterreichische Forschungsdokumentation" (the joint committee Austrian Research` |

Each cerif:project-abstract element is a one translation of the project abstract.

Translation is characterized by language, translation-type. For each cerif:project-abstract element you must create a

- Subelement cerif:proj_abs_language and its value must be code of the language of translation according to ISO 639 Codes. En for English, de for German
- Subelement cerif:proj_abs_trans_type which describes type of translation, O – original, H- human, M – Machine
- Subelement cerif:proj- abstract, describing the abstract (translation of abstract)

Documentation). The system is based on an Oracle database with currently more than 15.000 documents on research units and research activities. It is availabe online for database query. To ensure the acceptance of the AURIS database by the public, the project AURIS-MM will provide the following extensions to the AURIS service:</cerif:project.abstract>
        </cerif:project.project-abstract>
        </rdf:li>
</cerif:project.project-abstracts>
….
</cerif:project.project>

---

8 to provide information about keywords of the project you need to create an element cerif:project-keywords. It is an optional element, the project may but must not have keywords. If the project description has no keywords, this element must be omitted.

You can provide a number of keyworddescriptions for each translation of keywords.

Having created a cerif:project-keywords element, you must create a rdf:Bag subelement of it. And then for each keywords translation you must create a rdf:li element with a subelement cerif:project-keyword

Each cerif:project-keyword element is a one translation of project keywords.

Translation is characterized by language, translation-type. For each cerif:project-keywords element you must create a

- Subelement cerif:lang and its value must be code of the language of translation

<cerif:project.project-keywords>
  <rdf:Bag>
   <rdf:li>
    <cerif:project.project-keyword>
     <cerif:project.lang>en</cerif:project.lang>

<cerif:project.translation>H</cerif:project.translation>
      <cerif:project.keywords>multimedia; software; research information system</cerif:project.keywords>
<cerif:project.project-keyword>
    </rdf:li>
    <rdf:li>
    </rdf:li>
   </rdf:Bag>
  </cerif:project.project-keywords>

| according to ISO 639 Codes. En for English, de for German<br>■ Subelement cerif:translation which describes the type of translation, O – original, H-human, M – Machine<br>■ Subelement cerif:keywords, describing keywords (translation of keywords) | |
| --- | --- |

All data fields (RDF properties) isare divided into two types – mandatory and optional.  Mandatory RDF properties should be provided for any RDF resource description and its value must not be null zero???. If for mandatory data field is not provided or its value is null then all information about the resource is discarded and it will not be imported into the database. Also if there are other resources with mandatory relations to discarded  the resource, then those resources will also be discarded.

It is provided only one of tagging. But tags can be placed in the resource description in any order. In the future any tag can be used providing RDF Schema or OIL-DAML information about their semantic relations with CERIF-RDF tags.

## Step 3. Write  RDF data into files

To transfer information from data source to data target, it is required to transfer files with RDF description of data. There are several options how to transfer information

- All informationwhich you would like to transfer can be in one file. The order of resource description is arbitrary. But every description should be in the file only one time. Each description must have unique identified and it should be unique in the file. The file name must be in the format ORGANIZATIONNAME.DATE.ALL . Where the organization name is an acronym (or name, if acronym does not exist)  of data sourceorganization, according Austrian vocabulary of organizations. DATE is a date of file export in format DD.MM.YYYY. The date of file export from the database must be used, but not the date of transfer. ALL is a flag, saying that in this file all information is transferred.  As for example, a file of  full information from TUWIEN, generated 6 June 2001, would  have name TUWIEN.06.06.2001.ALL

- Each RDF description can be put in a separate file. So you haveone file for each project, organization, person. In each data transfer session for each object (project, person, organization) only one file must be generated.  Each resource in each file must have an identifier and it must be unique in the session. The file name must be in the format ORGANIZATIONNAME DATE.TYPE.IDENTIFIER. Where the organization name and date are the same this is described in the last paragraph. TYPE is a type of RDF description, it must be one of project, person, orgunit, equipment, result, publication, patent. IDENTIFIER is an identifier of an object, used in RDF description, As for example, file name for information about project E015-01-08, generated at TUWIEN 6 June 2001, must be named TUWIEN.06.06.2001.PROJECT. E015-01-08

All description of relations between objects must be put in every file describing this objects. It means, forexample, that a RDF description of the relation between project E015-01-08 and the person "Walter Niedermayer" must be putboth in  E015-01-08 as well as in"Walter Niedermayer" files.

An identifier once assigned to an object must be repeated in every data transfer session.

## Step 4.  Send files to AURIS

AURIS is supposed to provide access to actual research information in Austria. In this case it can attract new investors, collaborators to Austrian research. To make data actual, there are several good solutions

- To develop the Semantic Web solution, which enables the user to work with and search your data in your server anytime
- To have on the AURIS-MM server an actual copy of your data

The first solution is the final aim of the AURIS-MM system. But the second is easy to implement and it is a good test step to reach the final aim – the semantic web.

To make AURIS-MM containing actual information

1. each university and institution is **supposed** to send information about the full database once a year. It must be sent in a file with name ANNUAL.ORGANIZATION.YEAR.rdf. where ORGANIZATION is the organization acronym or name, year is a four-digit code of year. As example, annual report for Technical University of Vienna can be named ANNUAL.TUWIEN.2001.rdf

It is **desired** that each change of information about the project, orgunit or persons, information with a new description will be sent to AURIS. The file containing the new information must be named CHANGE.ORGANIZATION.TYPE.ID.DATE.rdf, where CHANGE is a flag that it is the actual new information, ORGANIZATION is a acronym or name of the data source organization, TYPE is a type of object, ID is an identifier of the object in the source database, DATE is a date of change (now the date when the information was sent). As for example, the file can be named CHANGE.TUWIEN.PROJECT.AURIS.01.05.2001.rdf

# How to encode all ERGO and AURIS data into CERIF RDF

A lot of organizations have experience in the generation of XML-SGML for the AURIS system. This experience is very useful for the new RDF generation. This paragraph provides guidelines how AURIS SGML data could be used to generate RDF description.

Information about the organization

| <RECORD> | For each record cerif:orgunit must be generated |
|---|---|
| <HRU>Niedermayer, Walter (Dipl.-Ing.) | |
| <KUG>Wissenschaftsinformation | |
| <KUG>FoDok-Austria | |
| <KUG>Forschungsinformationssystem | |
| <KUG>Datenbank Эber Forschung | |
| <KUG>Recherche von Forschungsinformationen | |
| <KUG>UniversitЙtsforschung: Informationen | |
| <KUG>Forschung-online | |
| <KUG>BroschЭren: spezielle Forschungsinformationen | |
| <KUG>Forschungsinformations-Service: Vermittlungsstelle | |
| <RUG>Wissenschaftsinformation - Forschungsdokumentation | |
| <RUE>Science Information - Research Documentation | |
| <KUE>Scientific information service | |
| <KUE>Research information system | |
| <KUE>FoDok | |
| <KUE>Database on research | |
| <KUE>University research | |
| | |
| <KUE>Research-online | |
| <KUE>Research information: services | |
| <KUE>Research documentation | |
| <DUG>Das Informationssystem ″FoDok″ der TU Wien - eine strukturierte Sammlung der Forschungsaktivitйten der Technischen Universitйt Wien - ist eine Жffentlich zugйngliche Service-Einrichtung am Auъeninstitut der TU Wien. Der Zugang zu den Informationen erfolgt hauptsйchlich Эber Recherchen in der Datenbank, Эber WWW online, oder mittels eigens entwickelter Recherche-Werkzeuge.<BR>In regelmйъigen Abstйnden (ca. alle 2 Jahre) wird der Datenbank-Inhalt als Buch (auch auf Diskette erhйltlich) herausgegeben. Auf Wunsch werden BroschЭren Эber spezielle Fachthemen erstellt. Die Forschungsinformationen dienen ebenso zur Vermittlung universitйrer Experten.<BR>Die Datenbank wird laufend aktualisiert. | |
| <DUE>″FoDok″, the research information system of the Vienna University of Technology is based on a structured database, which is accessible by public and which stores relevant information on the universities research activities.<BR>Access is guaranteed mainly by querying the database, either via WWW, in online mode, or by using retrieval software which is especially developed for that purpose.<BR>The contents of the database are published on a regular basis in intervals of about 2 years as book (and floppy disc). On demand special booklets on a specific theme can be created for specialists queries. And the research information is also used to ease the search for specialists on a specific theme.<BR>The database is updated constantly. | |
| <SEQ>Datenbank: Forschungsinformation | |
| <SRC>Andreas Krieger | |
| <UPD>2001-02-20 | |
| <CON> | |
| <STR>Guβhausstraβe 28 | |
| <PCD>A-1040 | |
| <TWN>Wien | |
| <TAC>+43 1 | |
| <TEL>58801 41522 | |
| <FAX>5054961 (58801 41599) | |
| <EML>walter@derpi.tuwien.ac.at | |
| <URL>http://derpi.tuwien.ac.at/walter | |
| <RCN>E015-01 | |
| <UNG>Technische Universität Wien | |
| <UNE>Vienna University of Technology | |
| <FAG>Dienstleistungseinrichtungen und Senatsinstitute | |
| <FAE>No Faculty | |
| <DEG>Außeninstitut | |
| <DEE>University Extension Centre | |
| </RECORD> | |

## Recommendations

### *Which of the above publishing formats / technologies - good tools are available for on-line publishing?*

- **JavaServer Pages™ (JSP™)** technology allows web developers and designers to rapidly develop and easily maintain, information-rich, dynamic web pages that leverage existing business systems. As part of the Java™ family, JSP technology enables rapid development of web-based applications that are platform independent. JavaServer Pages technology separates the user interface from content generation enabling designers to change the overall page layout without altering the underlying dynamic content. There are a number of free-of-charge implementaion of JSP. In AURIS-MM Tomcat (http://java.sun.com/products/jsp/tomcat/) server is used. It proved to be a good, easy-to use tools for developing internet applications
- **Active Server Pages$^T$ (ASP$^T$)** technology and language developed by Microsoft as a language for developing applications under Microsoft Internet Information Server platform. Now, not only MS IIS, but also Apache supports this technology
- **Oracle Application Server (OAS).** OAS is commercial Oracle software. Application can be created on a number of languages, such as Perl, PL/SQL. OAS is excellently integrated with other Oracle software products

### *- the published objects are machine readable*

To read automatically information published in AURIS-CERIF RDF/XML, XML or RDF parsers can be

used
We strongly advise you use DOM, not SAX parsers due to easy of programming
Comparison of XML parsers http://msdn.microsoft.com/library/periodic/period99/XMLpars.htm
Free XML software list  (http://www.oasis-open.org/cover/freeXMLSW980513.html)
In AURIS-MM we use Apache XM parses (Xerces) and it seems good, efficient software tools with little bugs

## RDF tools
List of RDF tools by Dave Beckett http://www.ilrt.bris.ac.uk/discovery/rdf/resources/#sec-apps,
Redland RDF application framework http://www.redland.opensource.ac.uk/
RDF parsers http://www.ilrt.bris.ac.uk/discovery/rdf-dev/roads/subject-listing/rdfparser.html
Semantic Web Community portal http://www.semanticweb.org/

## - hyperlinking objects is possible

In CERIF data model, describing research and academician data, there are a lot of relations between resources. As for example, there is a relationship of employment  between a person and an organization. To make a  hyperlink in an RDF document, you must use the "'resource=" notation.
To define that element person.orgunit is a hyperlink to the organization resource "Tuwien", it must be written
<person.orgunit resource="Tuwien"/>

## semantics can be retrieved by a future AURIS-MM agent
RDF is semantic. RDF allows not only to encode data, but to encode their meaning, make it possible to retrieve means of metadata elements ???by software agents
To define the meaning of  nRDF element, you can use the
-    RDF Schema  (http://www.w3.org/TR/rdf-schema)
-    OIL (http://www.ontoknowledge.org/oil/)
In AURIS-MM the RDF Schema is used for defining means of RDF elements when it is possible. To define the  means of an element in more complicated cases AURIS-MM uses OIL.
We use tools
-    http://img.cs.man.ac.uk/oil/   for editing of OIL ontology for  AURIS
-    http://www.cs.man.ac.uk/~horrocks/FaCT/ FaCT Description Logic Classifier   to reason over OIL

## internet search engines can find the pages and index them.
Internet engines such as www.Google.com, www.altavista.com, www.yahoo.com can find and index pages. To make your pages findable by these engines, the fastest way is to point engines on your pages. You can do it using
http://www.google.com/addurl.html
http://www.altavista.com/sites/search/addurl
We advice you to register you site in AURIS-MM MetaSearch facility
To add web search facilities to your site, we advice you to sue
MnogoSearch   enginer
http://www.mnogosearch.ru/

### - CERIF 2000 can be respected? What is the benefit of CERIF 2000?

The main benefits of CERIF-2000
- CERIF-2000 has being implemented on several CRIS (ERIS, Scottish Research, Slovakian CRIS, NIWI CRIS, etc) .
- There are now other accepted standards in Europe and US for research information, which cover such wide range of resources, important for CRIS
- Other standards or metadata formats can not completely describe what CERIF can. As example San Diego Supercomputer Center format for CV ([ftp://ftp.cordis.lu/pub/cris2000/docs/baru_fulltext.pdf](ftp://ftp.cordis.lu/pub/cris2000/docs/baru_fulltext.pdf)) can not describe
   - Facilities services of researchers
   - In which projects the researcher takes part in or took part in
   - Roles in expertise skills
   - Prizes and awards of researcher, academician and honorific titles
   - 

# Annexes. XML Schema, RDF Schema, Valuators

Technical issues

-- how to map your database into CERIF. Description of how to map with examples (AURIS-CERIF, FODOK-CERIF)
-- description of general schema mapping problems and what  to do to correct them
-- description of possible schematic conflicts and what to do to eliminate them
-- vocabulary mapping. what to do if your system uses vocabularies different   from AURIS
   -- how to map vocabularies
   -- how to provide AURIS team your vocabularies with mappings
-- how to generate RDF, using defined schema and vocabulary mapping from your database into CERIF
   -- general description of RDF tools
   -- description of our tool
-- how to check validity of descriptions
-- how to send data into AURIS
-- how to see entered data in AURIS tocheck them

Annexis
  Detail documented description of RDF schema
  Examples